\documentstyle[12pt,a4wide]{article}
\begin{document}
\input epsf
\newcommand{\be}{\begin{equation}}
\newcommand{\ee}{\end{equation}}
\newcommand{\pr}{\partial}
\newcommand{\ie}{{\it ie }}
\newcommand{\bphi}{\mbox{\boldmath $\phi$}}
\newcommand{\bpi}{\mbox{\boldmath $\pi$}}
\newcommand{\pauli}{\mbox{\boldmath $\tau$}}
\font\mybb=msbm10 at 11pt
\font\mybbb=msbm10 at 17pt
\def\bb#1{\hbox{\mybb#1}}
\def\bbb#1{\hbox{\mybbb#1}}
\def\bZ {\bb{Z}}
\def\bR {\bb{R}}
\def\bE {\bb{E}}
\def\bT {\bb{T}}
\def\bM {\bb{M}}
\def\bC {\bb{C}}
\def\bA {\bb{A}}
\def\bP {\bb{P}}
\def\e  {\epsilon}
\def\bbC {\bbb{C}}
\newcommand{\CP}{\bC \bP}
\newcommand{\CPP}{\bbC \bbb{P}}
\renewcommand{\theequation}{\arabic{section}.\arabic{equation}}
\newcommand{\news}{\setcounter{equation}{0}}
\newcommand{\I}{{\cal I}}
\newcommand{\HH}{\bb{H}^3_\kappa}
\newcommand{\bx}{{\bf x}}

\title{\vskip -70pt
\begin{flushright}
\end{flushright}\vskip 50pt
{\bf \large \bf SKYRMIONS, INSTANTONS, MASS AND CURVATURE}\\[30pt]
\author{Michael Atiyah$^{\ \dagger}$ and Paul Sutcliffe$^{\ \ddagger}$
\\[10pt]
\\{\normalsize $\dagger$ {\sl School of Mathematics,}}
\\{\normalsize {\sl University of Edinburgh,}}
\\{\normalsize {\sl King's Buildings, Edinburgh EH9 3JZ, U.K.}}
\\{\normalsize {\sl Email : atiyah@maths.ed.ac.uk}}\\
\\{\normalsize $\ddagger$  {\sl Institute of Mathematics,}}
\\{\normalsize {\sl University of Kent,}}\\
{\normalsize {\sl Canterbury, CT2 7NZ, U.K.}}\\
{\normalsize{\sl Email : P.M.Sutcliffe@kent.ac.uk}}\\}}
\date{November 2004}
\maketitle

\begin{abstract}
\noindent 
We show that Skyrmions with massless
pions in hyperbolic space provide a good approximation
to Skyrmions with massive pions in Euclidean space,
for a particular relationship between
the pion mass and the curvature of hyperbolic space. Using this result
we describe how a Skyrmion with massive pions in Euclidean space can be
approximated by the holonomy along circles of a Yang-Mills instanton.
This is a generalization of the approximation of Skyrmions by the
holonomy along lines of an instanton, which is only applicable
to massless pions.

\end{abstract}

\newpage\section{Introduction}\news\label{sec-intro} 
 
\ \quad The Skyrme model \cite{Sk} is a nonlinear theory of pions whose
topological soliton solutions  
are candidates for an effective description of nuclei, with an identification
between soliton and baryon numbers. A term can be included in the Skyrme
Lagrangian which gives the pions a mass, but as
the pion mass is relatively small this term is often neglected. 
For massless pions a Skyrmion has an algebraic 
asymptotic behaviour, but if
the pions are given a mass then the main effect is that the 
Skyrmion becomes exponentially localized.  Provided
the baryon number is low then a small pion mass has little qualitative effect
beyond changing the localization. However, a recent study \cite{BS7}
has shown that massive pions can have dramatic effects,
if either the baryon number or the pion mass is not small. 
For massless pions all the known minimal energy Skyrmions 
have a shell-like structure \cite{BS3} but for massive pions
some of these configurations are no longer bound states,
and there appear to be minimal energy Skyrmions which are 
finite chunks of the Skyrme crystal; this is an infinite triply periodic 
Skyrme field with a very low energy per baryon \cite{CJJVJ,KSh}. 
These results motivate a further study of Skyrmions with 
massive pions. 

In this letter we show that there is a surprising similarity 
between Skyrmions with massive pions in Euclidean space and 
Skyrmions with massless pions in hyperbolic space, with a relation
 between the pion mass and the curvature of hyperbolic space.
We map hyperbolic Skyrmions to Euclidean Skyrmions,
by identifying the hyperbolic and Euclidean radii, and find
that this gives very good approximations for a range of pion masses.

There are some very useful approximate methods for studying Skyrmions
with massless pions.
One approach is the rational map ansatz \cite{HMS}, where the 
angular dependence of the Skyrmion is determined by a rational map
between Riemann spheres. This approach is easily extended to the
case of massive pions, where again it provides good results \cite{BS7}, but it
can only describe Skyrmions which are shell-like. 
For massive pions some minimal energy
Skyrmions are not shell-like so an alternative approach is required.

Skyrmions with massless pions can be approximated by computing the 
holonomy of $SU(2)$ Yang-Mills instantons in Euclidean $\bb{R}^4$ along lines
parallel to the Euclidean time axis \cite{AM}. It seems likely that all
minimal energy Skyrmions can be approximated by the holonomy of a 
suitable instanton and a number of examples have been studied in detail
\cite{AM2,LM,SiSu,Su}. This approach is more general than the rational map 
ansatz, since it is not restricted to shell-like configurations.
Furthermore, it is known \cite{MS} that there is an instanton on $\bb{T}^4$
whose holonomy gives an approximation to the Skyrme crystal, so it seems
likely that crystal chunks could also be approximated by instanton 
holonomies.

Instanton generated Skyrme fields have the correct algebraic decay
for Skyrmions with massless pions, but they can not
be used to approximate Skyrmions with massive pions, since the decay
is too slow to yield finite energy. It would therefore be desirable 
to have a generalization of the instanton holonomy method that could
produce Skyrme fields with the exponential decay appropriate for
massive pions. Such a geneneralization is presented here. 
Computing the holonomy of an instanton along particular circles in 
$\bb{R}^4$ yields a Skyrme field in hyperbolic 3-space, and applying
our mapping to this hyperbolic Skyrme field produces
a Skyrme field in Euclidean space with exponential decay. 
In the limit of massless pions this procedure coincides
with the usual instanton holonomy along lines, but it is applicable
to any value of the pion mass. We apply this scheme explicitly to
the single Skyrmion for a range of pion masses and find that the
approximation actually improves with increasing pion mass, so that the
error is always less than the usual case with massless pions.

\section{Hyperbolic Skyrmions and the pion mass}\news\label{sec-hyp} 
 
\ \quad The static energy of the $SU(2)$-valued Skyrme field $U({\bf x})$
on a 3-dimensional Riemannian manifold $M$ with metric 
$ds^2=g_{ij}dx^idx^j$
 is given by
\be
E=\frac{1}{12\pi^2}\int \left\{-{1 \over 2}\mbox{Tr}(R_iR^i)-{1 \over 16}
\mbox{Tr}([R_i,R_j][R^i,R^j])+m^2\mbox{Tr}(1-U)\right\} 
\sqrt{g} \, d^3x\,,
\label{skyenergy}
\ee
where $R_i=(\partial_i U)U^\dagger$ is the $su(2)$-valued current,
$g$ denotes the determinant of the metric and $m$ is the pion mass
parameter.

It is easy to see that $m$ is the (tree-level) mass of the pions 
by making contact with the nonlinear pion theory via
$
U=\sigma +i\bpi\cdot\pauli,$
where $\bpi=(\pi_1,\pi_2,\pi_3)$ is the triplet of pion fields,
$\pauli$ denotes the triplet of Pauli matrices
 and $\sigma$ is determined by the constraint $\sigma^2+\bpi\cdot\bpi=1.$

In this paper we shall be concerned with two choices for $M.$
The first choice is simply Euclidean space $M=\bb{R}^3,$ where 
we are mainly concerned with the properties of Skyrmions for varying values 
of the pion mass $m.$ The Skyrme model has energy and length units which have
 been scaled out in the expression (\ref{skyenergy}). These energy and
length units must be fixed by comparison with experiment, and then $m$
denotes the pion mass in these units. The standard approach to fixing these
units involves fitting the masses of the proton and delta resonance \cite{ANW}
 and this leads to a value \cite{AN} of $m=0.526.$ However, other 
approaches to fixing the units are possible and these would produce 
different values of $m$ corresponding to the physical pion mass, so it
is worth investigating the properties of Skyrmions as a function of $m.$

The second choice we shall discuss is where $M=\HH,$ hyperbolic 3-space
of constant negative curvature $-\kappa^2.$ In this case we restrict
to massless pions ($m=0$) and consider the properties of Skyrmions
for varying values of the curvature $-\kappa^2.$

In the following we shall make a surprising connection between these
two apparently quite different situations.


Let us first consider the Euclidean case.
The single Skyrmion has the hedgehog form
\be
U=\mbox{exp}(if(r)\widehat\bx\cdot\pauli)
\ee
where $f(r)$ is the radial profile function with boundary conditions
$f(0)=2\pi$ and $f(\infty)=0.$ The energy of this field is given by 
 \be
E=\frac{1}{3\pi}\int \bigg( r^2f'^2+2(f'^2+1)\sin^2 f+\frac{\sin^4
f}{r^2}+2m^2r^2(1-\cos f)\bigg) \ dr.
\label{euclideanenergy}
\ee 
The energy minimizing profile function has the asymptotic exponential
decay 
\be
f\sim \frac{A}{r}e^{-mr}
\label{edecay}
\ee
but in the massless pion limit ($m=0$) this is replaced by the
algebraic form $f\sim Ar^{-2}.$

\begin{figure}[ht]
\begin{center}
\leavevmode
\vskip -3.5cm
\epsfxsize=16cm\epsffile{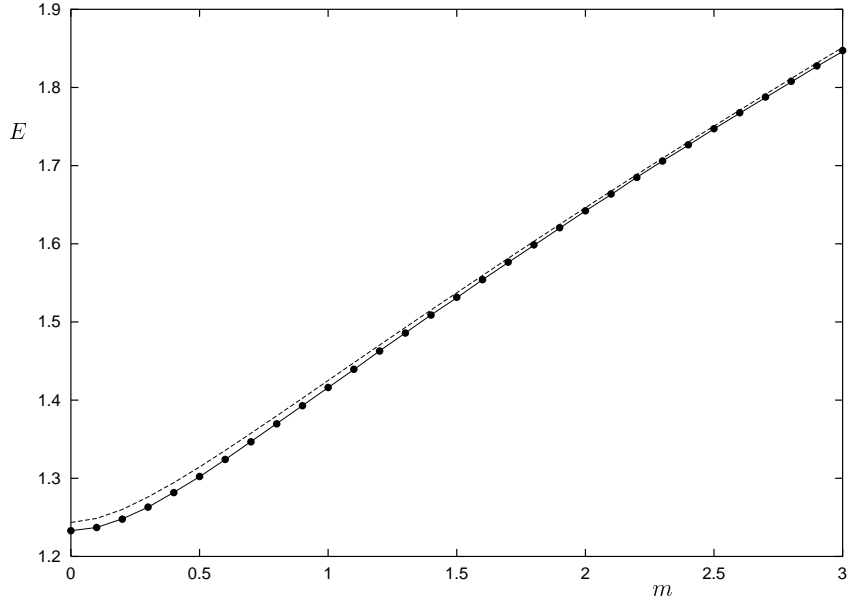}
\vskip -11.5cm
\caption{The Skyrmion energy as a function of the pion mass;
exact result (solid curve), hyperbolic approximation (circles),
instanton approximation (dashed curve).}
\label{fig-e1}
\vskip 0cm
\end{center}
\end{figure}

A numerical solution of the ordinary differential equation which
follows from (\ref{euclideanenergy}) allows the Skyrmion 
energy to be computed as a function of the pion mass and this is 
displayed as the solid curve in Fig.~\ref{fig-e1}.
\begin{figure}[ht]
\begin{center}
\leavevmode
\vskip -3.5cm
\epsfxsize=16cm\epsffile{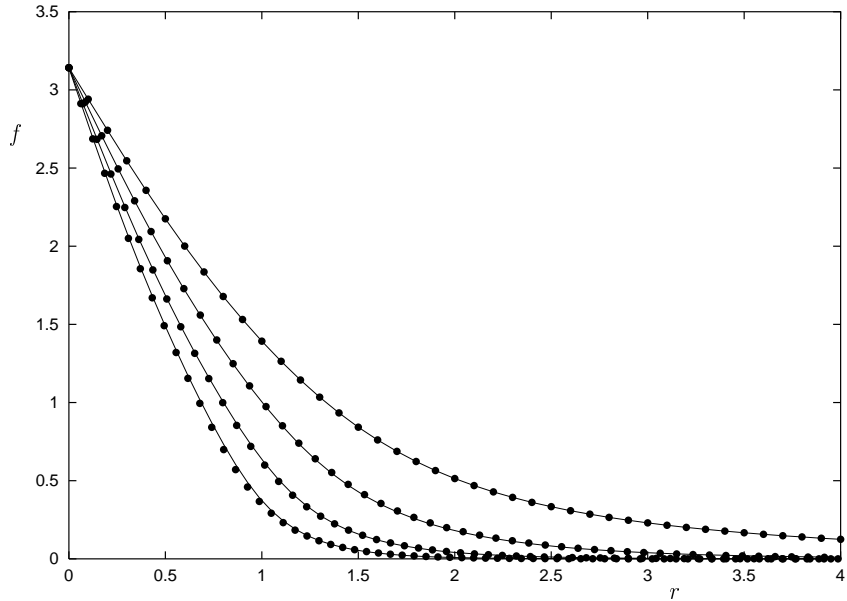}
\vskip -11.5cm
\caption{Profile functions for pion masses $m=0,1,2,3$; 
exact result (solid curves) and
hyperbolic approximation (circles). The curves are more localized for
larger $m.$ }
\label{fig-pro1}
\vskip 0cm
\end{center}
\end{figure}
The profile functions for $m=0,1,2,3$ are shown as the solid curves
in Fig.~\ref{fig-pro1}, demonstrating the increased localization 
with increasing pion mass.

Now consider hyperbolic Skyrmions with massless pions.
In spherical coordinates the metric of $\HH$ takes the form
\be
ds^2(\HH)=d\rho^2+\frac{\mbox{sinh}^2\kappa\rho}{\kappa^2}
(d\theta^2+\sin^2\theta\,d\phi^2)
\label{metrichyp}
\ee
where $\rho$ is the hyperbolic radius and the curvature of hyperbolic
space is $-\kappa^2.$ Note that in the limit as $\kappa\rightarrow 0$
the Euclidean metric is recovered with $\rho=r$ the usual radial coordinate.

The hedgehog form 
\be
U=\mbox{exp}(iF(\rho)\widehat\bx\cdot\pauli)
\label{hhh}
\ee
gives the energy of the hyperbolic Skyrmion to be
 \be
E=\frac{1}{3\pi}\int \bigg(
F'^2\frac{\mbox{sinh}^2\kappa\rho}{\kappa^2}
+2(F'^2+1)\sin^2 F+\frac{\kappa^2 \sin^4 F}{\mbox{sinh}^2\kappa\rho}\bigg)
 \ d\rho.
\label{hyperbolicenergy}
\ee 
The energy minimizing profile function again has an asymptotic exponential
decay
\be
F\sim Ae^{-2\kappa\rho}.
\label{hdecay}
\ee
Although the Skyrme energy functions 
in Euclidean space with massive pions and hyperbolic space with massless
pions are quite different they coincide in the limits
$m=0$ and $\kappa\rightarrow 0,$ where the massless pion Euclidean 
Skyrmion is recovered. Both modifications away from this limit yield
exponentially localized solutions. Furthermore, close to the
origin of hyperbolic space it increasingly resembles Euclidean space so
the influence of both curvature and the pion mass should be greatest
outside the Skyrmions core.
Thus it seems reasonable to ask if there
are any similarities between these two systems.

To compare a Skyrme field in hyperbolic space to one in Euclidean space
we simply map the hyperbolic radius $\rho$ to the 
Euclidean radius $r$ (note that this requires a choice of origin
and so breaks the translation invariance of $\bb{R}^3$).
 In the case of a single Skyrmion this means
comparing the profile functions $f(r)$ and $F(\rho=r)$. It is
interesting to see if there is a relationship between the two deformation
parameters $m$ and $\kappa$ so that these profile functions are
similar. 

\begin{figure}[ht]
\begin{center}
\leavevmode
\vskip -3.5cm
\epsfxsize=16cm\epsffile{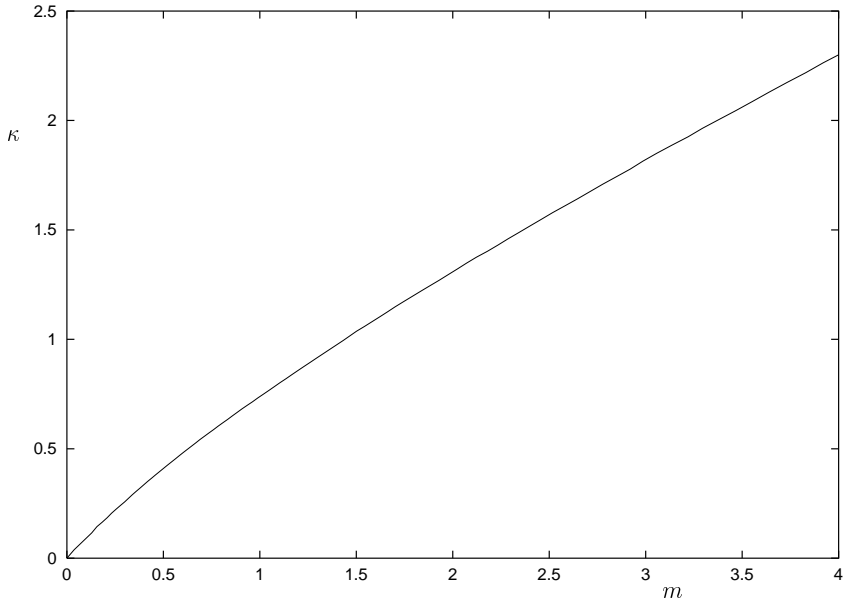}
\vskip -11.5cm
\caption{$\kappa$ as a function of $m,$ obtained by equating the size
of a Skyrmion.}
\label{fig-kvm}
\vskip 0cm
\end{center}
\end{figure}

Comparing the leading order of the asymptotic decays (\ref{edecay}) and 
(\ref{hdecay}) suggests the simple
linear relationship $\kappa=m/2.$ However, since this result is obtained
using only the large radius behaviour, where the difference between
Euclidean and hyperbolic space is greatest, then we expect this simple
linear relationship to be only a rough guide. 
As the size of a Skyrmion depends both on the pion mass and the curvature of
 hyperbolic
space then a better method for 
determining a possible relationship between $\kappa$ and $m$ is to
fix the size of the Skyrmion to be the same in both cases. The
size of a Skyrmion is defined to be the radius at which the profile
function takes the value $\pi/2,$ so that the $\sigma$ field vanishes.
Thus for each value of $m$ we compute $r_*$ such that $f(r_*)=\pi/2$
and then determine $\kappa$ by the requirement that $F(r_*)=\pi/2.$
The result is the relationship $\kappa(m)$ plotted in Fig.~\ref{fig-kvm},
which is reasonably close to the naive linear estimate $\kappa=m/2.$

Given the relationship $\kappa(m)$ we can now investigate whether
the Skyrmion in hyperbolic space of curvature $-(\kappa(m))^2$
with massless pions is a good approximation to the Skyrmion in Euclidean
space with pion mass $m.$ To do this we take the hyperbolic Skyrmion,
identify hyperbolic and Euclidean radii by making the replacement
$\rho\mapsto r,$ and compute the Euclidean energy of this Skyrme
field, to determine its excess over the exact energy. The energy
of the hyperbolic approximation is displayed as
the circles in Fig.~\ref{fig-e1} for a range of $m.$
It can be seen from this figure that the hyperbolic approximation
is remarkably close to the true solution, and in fact the excess energy is
always less than $0.1\%$ over this range of pion masses.
In Fig.~\ref{fig-pro1} the circles display the hyperbolic profile function 
$F(\rho=r)$ for $\kappa(m)$ with $m=0,1,2,3,$ for comparison with the
true profile functions $f(r)$ (solid curves). Again this demonstrates
that the hyperbolic approximation is remarkably accurate.

So far our study has been limited to the single Skyrmion, but now we turn
to multi-Skyrmions. Ideally, the comparison of hyperbolic and Euclidean
multi-Skyrmions would involve full numerical simulations of both  
nonlinear field theories, but this is computationally expensive. As an
alternative we shall investigate multi-Skyrmions within the framework of 
the rational map ansatz. For Euclidean Skyrmions with massive pions
the rational map approximation has already been compared to full field
simulations \cite{BS7} and found to be an excellent approximation for
shell-like configurations for the whole range of pion masses studied.
At least for baryon numbers $B\le 4$ it appears that shell-like
Skyrmions survive as the minimal energy configurations even for quite
large pion masses, so the rational map approach is justified.

We can treat both Euclidean and hyperbolic Skyrmions simultaneously
by first considering the Skyrme model on hyperbolic space with a pion mass
$\widetilde m$
and then taking the limit $\kappa\rightarrow 0$ with $\widetilde m=m$ 
  to recover the
Euclidean case or the limit $\widetilde m=0$ with $\kappa=\kappa(m)$ 
to obtain the hyperbolic massless pion case.

The rational map ansatz \cite{HMS} constructs a Skyrme field with 
baryon number $B$ from a degree
$B$ rational map between Riemann spheres. Although this
ansatz does not give exact multi-Skyrmion solutions of the static
 Skyrme equations,
it produces approximations which (at least in the Euclidean case)
 have energies only a few percent
above the numerically computed solutions. Furthermore, the
 approximation of the angular dependence involved in this approach
should produce similar errors in both Euclidean and hyperbolic space,
so we can use the rational map determined energies to compare these two
cases without worrying about the small common error.

To present the rational map ansatz we first introduce the Riemann sphere
coordinate $z=e^{i\phi}\tan(\theta/2),$ and let 
$R(z)$ be a degree $B$ rational map between Riemann spheres,
that is, $R=p/q$ where $p$ and $q$ are polynomials in $z$ such that
$\max[\mbox{deg}(p),\mbox{deg}(q)]=B$,  and $p$ and $q$ have no common
factors.  Given such a rational map the ansatz for the Skyrme field 
in hyperbolic space is
\be  U(\rho,z)=\exp\bigg[\frac{iF(\rho)}{1+\vert R\vert^2} \pmatrix{1-\vert
R\vert^2& 2\bar R\cr 2R & \vert R\vert^2-1\cr}\bigg]\,,
\label{rma}
\ee where $F(\rho)$ is a real profile function satisfying the  boundary
conditions $F(0)=\pi$ and $F(\infty)=0.$ This profile function 
is determined by minimization of the energy of the field (\ref{rma}) given a
particular rational map $R$. 

Substitution of the ansatz (\ref{rma}) into the hyperbolic Skyrme
energy results in the following expression 
 \be
E=\frac{1}{3\pi}\int \bigg(
F'^2\frac{\mbox{sinh}^2\kappa\rho}{\kappa^2}
+2B(F'^2+1)\sin^2 F+\I\frac{\kappa^2 \sin^4 F}{\mbox{sinh}^2\kappa\rho}
+2\widetilde m^2\frac{\mbox{sinh}^2\kappa\rho}{\kappa^2}(1-\cos F)\bigg)
 \ d\rho
\label{rmaenergy}
\ee 
where $\I$ denotes the integral 
\be \I=\frac{1}{4\pi}\int \bigg(
\frac{1+\vert z\vert^2}{1+\vert R\vert^2}
\bigg\vert\frac{dR}{dz}\bigg\vert\bigg)^4 \frac{2i \  dz  d\bar z
}{(1+\vert z\vert^2)^2}\,.
\label{i}
\ee 
For $B=2,3,4$ the rational maps which minimize $\I$ are given by
\be
R=z^2 ,\quad
R=\frac{z^3-\sqrt{3}iz}{\sqrt{3}iz^2-1},\quad
R=\frac{z^4+2\sqrt{3}iz^2+1}{z^4-2\sqrt{3}iz^2+1}.
\label{234maps}
\ee 
\begin{figure}[ht]
\begin{center}
\leavevmode
\vskip -3.5cm
\epsfxsize=16cm\epsffile{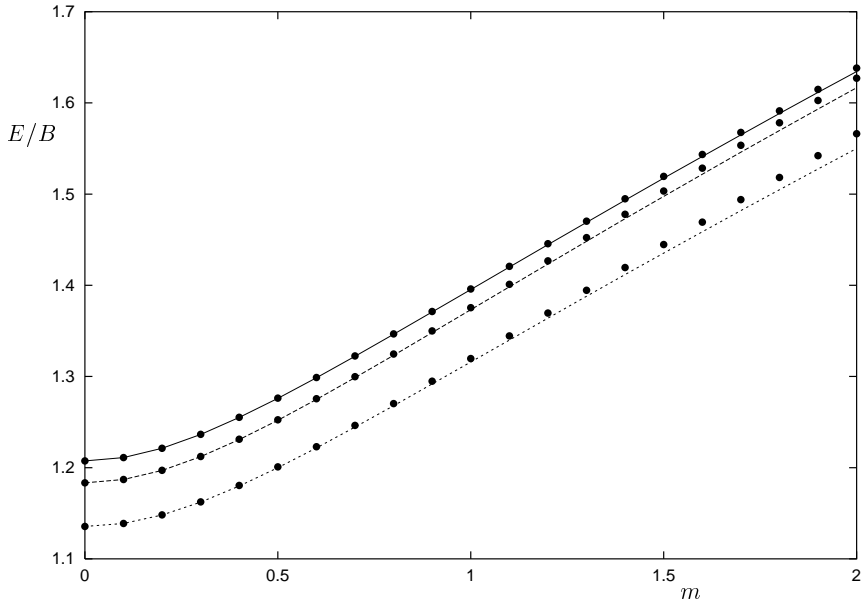}
\vskip -11.5cm
\caption{The energy per baryon $E/B$ as a function of the pion mass
$m$ for $B=2$ (solid curve), $B=3$ (dashed curve), $B=4$ (dotted curve).
The circles denote the energies of the hyperbolic approximations.}
\label{fig-e234}
\vskip 0cm
\end{center}
\end{figure}

Using these maps we first compute the Euclidean energy 
($\kappa\rightarrow 0$) with pion mass $\widetilde m=m.$ 
The results are presented in Fig.~\ref{fig-e234} where we plot
 the energy per baryon $E/B$ for $B=2$ (solid curve), 
$B=3$ (dashed curve), $B=4$ (dotted curve). Next we use the same 
rational maps and compute the
hyperbolic Skyrme fields for massless pions ($\widetilde m=0$) with
$\kappa=\kappa(m)$, and calculate their Euclidean energy with pion mass $m.$
The energies of these hyperbolic Skyrmion approximations are shown as
the circles in Fig.~\ref{fig-e234}. We see from this figure that the
hyperbolic Skyrme fields are also a good approximation to multi-Skyrmions,
though for increasing pion mass the errors are greater for the larger
baryon numbers. Note that at a pion mass around the standard one used in
the literature, $m=0.526,$ the errors are very small even for the
larger baryon numbers. 

In this section we have demonstrated that hyperbolic Skyrmions with
massless pions provide a good approximation to Euclidean Skyrmions
with massive pions. However, hyperbolic Skyrmions are at least as
difficult to compute as Euclidean Skyrmions, so it might appear
that this connection is not very useful. There are two main reasons why
the relation with hyperbolic Skyrmions is of interest.

 The first reason is that
some insight into the more complicated effects of the pion mass, such
as the emergence of minimal energy crystal chunks, might be found
by studying the geometry of hyperbolic space. For example, the
infinite curvature limit may yield some simplifications as it
does in the study of hyperbolic monopoles \cite{A}. Also, the extra
term in the Skyrme Lagrangian that gives the pions a mass is only
really determined upto quadratic order in the pion fields. The term
that is generally used is just the simplest possibility,
 and therefore the most natural from one point of view, but the
correspondence with hyperbolic Skyrmions may suggest an alternative
choice motivated by the geometry. 

The second reason is a more practical one, in that it will allow
us to approximate Euclidean Skyrmions with massive pions in terms of
the holonomy of Yang-Mills instantons. This is explained in the 
following section.

\section{Instanton holonomies}\news\label{sec-inst} 
Let us first recall how Euclidean Skyrmions with massless pions
can be approximated using Yang-Mills instantons. This 
scheme was introduced in \cite{AM}
and involves computing the holonomy of $SU(2)$ instantons
in Euclidean $\bb{R}^4$
along lines parallel to the $x_4$-axis. Explicitly, the prescription
for the Skyrme field is to take
\be
U({\bf x})={\cal P}\exp\bigg(\int_{-\infty}^{\infty} A_4({\bf
x},x_4)\, dx_4\bigg)
\label{skyfromins}
\ee
where ${\cal P}$ denotes path ordering and $A_\mu$ is the gauge potential of
a Yang-Mills instanton in $\bb{R}^4$, and where
$x=({\bf x},x_4)=(x_1,x_2,x_3,x_4)$.
This holonomy is really along a closed loop in $S^4$, and is almost gauge
invariant. The only effect of a gauge transformation $g(x)$ is to
conjugate $U(\bx)$ by a fixed element $g(\infty),$ and this 
simply corresponds to an
isospin rotation of the Skyrme field. 

If $A_\mu$ is the field of a charge $N$ instanton then 
it follows from general topological
considerations that the resulting Skyrme field has baryon number $B=N$.
The construction yields
an $(8N-1)$-dimensional family of Skyrme fields from the $8N$-dimensional
moduli space of charge $N$ instantons, and although such
fields are never exact solutions of the Skyrme
equation, some are good approximations to minimal energy
Skyrmions and other important field configurations \cite{AM2,LM,SiSu,Su}.

As the simplest example, consider the charge one instanton centred at
the origin with width $\lambda.$ This is given by the 
't Hooft ansatz 
\be
A_\mu=\frac{i}{2}\sigma_{\mu\nu}\partial_\nu\log\zeta\,
\label{tH}
\ee
where 
\be
\zeta=1+\frac{\lambda^2}{|x|^2}
\label{onepole}
\ee
and $\sigma_{\mu\nu}$ is the anti-symmetric anti-self-dual tensor
defined by $\sigma_{i4}=\tau_i, \ \sigma_{ij}=\varepsilon_{ijk}\tau_k.$ 

The holonomy of this instanton
generates a Skyrme field of the hedgehog form with a profile function given by
\be
f(r)=\pi\left[1-\left(1+{\lambda^2\over r^2}\right)^{-1/2}\right] \,.
\label{insprofile}
\ee
Instantons are scale invariant, so the parameter $\lambda$ is arbitrary and can
be chosen to minimize the energy of the resulting Skyrme field.
The appropriate value of the scale is $\lambda^2=2.11$, and then the energy is
$E=1.243$, which is only $1\%$ above that of the true Skyrmion solution.

The instanton approximation has been very useful for studying
massless pion Skyrmions, so it would be useful if a similar approach was
available in the massive pion case. However, note that the asymptotic 
decay of the instanton generated profile function is $f\sim \lambda^2/(2r^2)$
and this has the correct form for massless pions, but for massive pions this
algebraic decay produces infinite energy. Thus this form of the instanton
holonomy method is not applicable to massive pions.

Skyrmions on hyperbolic space can be approximated by 
instanton holonomies along circles in $\bb{R}^4$ \cite{MSam}.
We can combine this approach with the results of the previous section,
relating hyperbolic Skyrmions to Euclidean Skyrmions with massive pions,
to obtain a generalization of the instanton approximation that is
valid for all pion masses. The details are as follows.

The starting point is to observe that there is a conformal equivalence
between $\bb{R}^4-\bb{R}^2$ and $\HH\times S^1_{\kappa^{-1}},$ where
$S^1_{\kappa^{-1}}$ denotes the circle of radius ${\kappa^{-1}}.$
One way to see this explicitly is to introduce toroidal coordinates
$(\rho,\theta,\phi,\chi)$ on $\bb{R}^4$ given by
\be
x=\frac{1}{\cosh({\kappa \rho})+\cos\chi}
(\sinh({\kappa \rho})\sin\theta\cos\phi,
\sinh({\kappa \rho})\sin\theta\sin\phi,
\sinh({\kappa \rho})\cos\theta,
\sin\chi)
\ee
where $\kappa$ is a real parameter.
It is then easy to check that the metric on $\bb{R}^4$ becomes
\be
ds^2(\bb{R}^4)=dx_1^2+dx_2^2+dx_3^2+dx_4^2
=\frac{(ds^2(\HH)+\kappa^{-2}d\chi^2)}{(\cosh({\kappa \rho})+\cos\chi)^2}
\ee
where $ds^2(\HH)$ is the metric on hyperbolic 3-space with spherical
coordinates $(\rho,\theta,\phi)$ as given in equation (\ref{metrichyp}).
Dropping the conformal factor in the final expression above we obtain
hyperbolic 3-space and a circle of radius $\kappa^{-1}$ with coordinate
$\chi.$

To obtain a Skyrme field in hyperbolic space the instanton holonomy
is computed along the circles parametrized by $\chi,$ that is
\be
U(\rho,\theta,\phi)={\cal P}\exp\bigg(\int_{0}^{2\pi} A_\chi\, d\chi\bigg)
\label{hskyfromins}
\ee
where $A_\chi$ is the component of the gauge potential associated with
the coordinate $\chi.$ 

As mentioned earlier, the construction of Skyrme fields on $\bb{R}^3$
from instanton holonomies along lines is essentially gauge invariant,
due to the fact that all the lines go through the point at infinity.
However, the circles whose
holonomy we use to generate hyperbolic Skyrme fields have no points in common,
so the holonomy is only defined up to conjugation unless we fix a
gauge on the whole of hyperbolic space.  
Explicitly, under a gauge transformation $g(\rho,\theta,\phi,\chi)$
the holonomy (\ref{hskyfromins}) transforms as
\be 
U(\rho,\theta,\phi)\mapsto g^{-1}(\rho,\theta,\phi,0)
U(\rho,\theta,\phi)g(\rho,\theta,\phi,0)
\ee
which is a local isospin rotation.

There is a rather natural resolution to this problem, namely 
we fix the gauge by requiring the radial gauge $A_\rho=0.$
This requires a choice of origin but this symmetry breaking choice
 is already needed in order to apply our mapping from hyperbolic Skyrmions
to Euclidean Skyrmions. 
 The upshot is that the Skyrme fields from instantons construction works with
full symmetry for the Euclidean case but only with rotational symmetry 
(around an origin) for the hyperbolic case. This is quite natural since
hyperbolic space has no translations.

Let us consider the simplest case of a single Skyrmion.
For the charge one instanton in the 't Hooft gauge (\ref{tH}) then
$A_\chi$ has the form $A_\chi=a_\chi(\rho,\chi) \widehat\bx\cdot\pauli,$ where
$a_\chi(\rho,\chi)$ is independent of $\theta$ and $\phi.$ 
In this gauge $A_\rho$ is non-zero but it has the form
$A_\rho=a_\rho(\rho,\chi) \widehat\bx\cdot\pauli,$ where
again $a_\rho(\rho,\chi)$ is independent of $\theta$ and $\phi.$ 
Since $A_\chi$ and $A_\rho$ both have the same direction in the Lie algebra,
which furthermore is independent of both $\rho$ and $\chi,$ then the
gauge transformation to the radial gauge $A_\rho=0$ has no effect on the
holonomy (\ref{hskyfromins}) and therefore does not need to be 
found explicitly.

Computing the charge one instanton holonomy yields a hyperbolic Skyrme
 field of the
hedgehog form (\ref{hhh}) with profile function \cite{MSam} 
\be
F(\rho)=\pi\bigg(1-\sqrt{\frac{\sinh^2(\kappa\rho)}
{\kappa^2\beta^2+\sinh^2(\kappa\rho)}}\bigg)
\label{hi}
\ee
where $\beta$ is related to the instanton scale $\lambda$ via
$\beta={2\lambda\kappa^{-1}}{/(1+\lambda^2).}$

Note that in the zero curvature limit, $\kappa\rightarrow 0$ with $\rho=r,$ 
the profile function (\ref{hi}) reverts to the Euclidean form 
(\ref{insprofile}). The fact that in the zero curvature limit 
the hyperbolic Skyrme field obtained from
the holonomy along circles reverts to the Euclidean Skyrme field constructed
from the holonomy along lines is a general feature, 
and corresponds to the fact that the holonomy is computed along circles
of radius $\kappa^{-1}$ which degenerate to lines.

For non-zero curvature the instanton generated hyperbolic Skyrme field
has an exponential decay, as can be seen explicitly in the case of the 
single Skyrmion profile function (\ref{hi}). 
Applying the mapping from hyperbolic to Euclidean 
space, by identifying radial coordinates, therefore produces
instanton generated Euclidean Skyrme fields which can be used to approximate
Skyrmions with massive pions.

Considering the single Skyrmion in detail, we take the instanton
generated Skyrme field with profile function (\ref{hi}) and make
the replacement $\rho\mapsto r.$ We then compute the Euclidean energy
of this approximation as a function of the pion mass, 
minimizing over the instanton scale, and with $\kappa$ determined
by the pion mass using the relation displayed in Fig.~\ref{fig-kvm}.
The result is displayed as the dashed curve in Fig.~\ref{fig-e1}.
For zero pion mass the result of ref.\cite{AM} is reproduced, with
the approximation exceeding the true energy by around $1\%,$ but as 
the pion mass increases it can be seen from  Fig.~\ref{fig-e1} that the
dashed curve approaches the solid curve (the true energy), showing that
the approximation actually improves with increasing pion mass, at least for the
range of values considered.

\begin{figure}[ht]
\begin{center}
\leavevmode
\vskip -3.5cm
\epsfxsize=16cm\epsffile{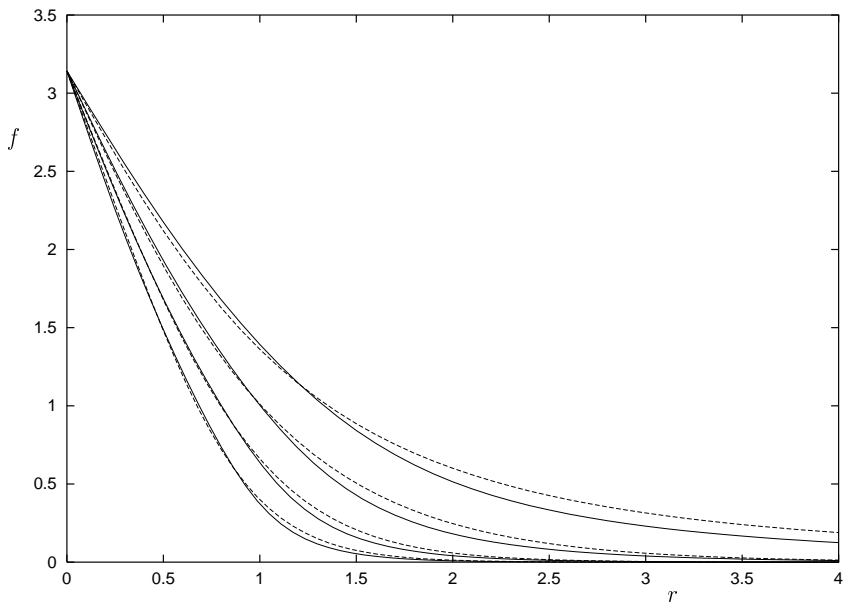}
\vskip -11.5cm
\caption{Profile functions for pion masses $m=0,1,2,3;$
exact results (solid curves) and instanton approximation (dashed curves).
The curves are more localized for larger $m.$}
\label{fig-promi}
\vskip 0cm
\end{center}
\end{figure}

As further indication that the instanton approximation improves with
increasing pion mass we compare in Fig.~\ref{fig-promi} the true profile
 functions (solid curves) and the instanton generated profile functions
 (dashed curves) for pion masses $m=0,1,2,3.$ The more localized 
curves correspond to larger values of $m.$ It is evident from this figure
that the instanton generated fields approximate
the exact solutions with an improved accuracy compared to the massless 
pion case. The instanton approximation for massless pions
is itself quite accurate and has proved useful for studying Skyrmions
 in a number of ways, so we expect that our modified instanton
proposal should be useful in the study of Skyrmions with massive pions. 

\section{Conclusion}\news\label{sec-con} 
We have shown that, for a particular relationship between the pion mass
and the curvature of hyperbolic space, there is a surprising
similarity between Euclidean Skyrmions with massive pions and
hyperbolic Skyrmions with massless pions. The latter provide very
good approximations to the former and we have shown how this can
be used to approximate Euclidean Skyrmions with a pion mass 
by the holonomy along circles of Yang-Mills instantons.

The connection between Euclidean and hyperbolic Skyrmions suggests
that it might be interesting to investigate hyperbolic Skyrmions in
more detail. Particularly interesting aspects include the behaviour 
of Skyrmions as the curvature tends to infinity, and the interpretation
of curvature as a Euclidean pion mass, which may lead to a more geometrically
natural mass term.

Given that the instanton holonomy approximation can now be applied
to Skyrmions with massive pions it is of interest to determine 
the instantons which are relevant in this context. In particular,
the results of ref.\cite{BS7} suggest that the structure and symmetries of 
some minimal energy Skyrmions will be quite different from the massless
pion situation, and it might be possible to understand this in terms
of instantons. For example, there is evidence that for massive pions 
the minimal energy charge 32 Skyrmion is a crystal chunk with cubic symmetry.
It would therefore be interesting to know if a charge 32 instanton
exists whose holonomy is a good approximation to this Skyrmion. One way
to approach this problem is to attempt to construct all charge 32 instantons  
with cubic symmetry and then investigate whether any of these have the
correct properties required of a crystal chunk. However, for such a large
charge there is likely to be a fairly big moduli space of such symmetric
solutions, so it may be necessary to first determine a good method for
identifying instantons which resemble crystal chunks.

Crystal chunks are not only interesting for instantons and Skyrmions,
but also for monopoles. It is known that there are many similarities
between monopoles and Skyrmions (see eg. ref.\cite{book}) and circle
invariant instantons can be identified with hyperbolic monopoles 
\cite{At}, so it is natural to wonder whether monopole chunks exist
in some form, even though there is no infinite monopole crystal. 
There is a diffeomorphism, which preserves rotational symmetry,
  between the moduli space of charge $N$ $SU(2)$ monopoles and 
the moduli space of 
degree $N$ rational maps between Riemann spheres \cite{Ja}. 
It is fairly easy to construct a two-parameter family of degree
32 rational maps with cubic symmetry, using the Klein polynomials
of the tetrahedron, but again it is not easy to see if any member
of this family corresponds to a charge 32 monopole which resembles 
the crystal chunk of the charge 32 Skyrmion. As in the instanton
situation it would be very useful if a property of this data
could be identified that would distinguish chunk-like monopoles
(if they exist) from those which are shell-like.

\end{document}